\documentclass{JHEP}
\usepackage{epsfig}
\usepackage[active]{srcltx}

\newcommand{\fref}[1]{Figure~\ref{#1}}
\newcommand{\cref}[1]{Chapter~\ref{#1}}
\newcommand{\beq}{\begin{equation}}
\newcommand{\eeq}{\end{equation}}
\newcommand{\ba}{\begin{array}}
\newcommand{\ea}{\end{array}}
\newcommand{\bcenter}{\begin{center}}
\newcommand{\ecenter}{\end{center}}

\def\IB{\relax\hbox{$\inbar\kern-.3em{\rm B}$}}
\def\IC{\relax\hbox{$\inbar\kern-.3em{\rm C}$}}
\def\ID{\relax\hbox{$\inbar\kern-.3em{\rm D}$}}
\def\IE{\relax\hbox{$\inbar\kern-.3em{\rm E}$}}
\def\IF{\relax\hbox{$\inbar\kern-.3em{\rm F}$}}
\def\IG{\relax\hbox{$\inbar\kern-.3em{\rm G}$}}
\def\IGa{\relax\hbox{${\rm I}\kern-.18em\Gamma$}}
\def\IH{\relax{\rm I\kern-.18em H}}
\def\IK{\relax{\rm I\kern-.18em K}}
\def\IL{\relax{\rm I\kern-.18em L}}
\def\IP{\relax{\rm I\kern-.18em P}}
\def\IR{\relax{\rm I\kern-.18em R}}
\def\IZ{\relax\ifmmode\mathchoice
{\hbox{\cmss Z\kern-.4em Z}}{\hbox{\cmss Z\kern-.4em Z}}
{\lower.9pt\hbox{\cmsss Z\kern-.4em Z}}
{\lower1.2pt\hbox{\cmsss Z\kern-.4em Z}}\else{\cmss Z\kern-.4em Z}\fi}
\def\II{\relax{\rm I\kern-.18em I}}

\def\sCC{{\kern 0.27em\vrule height1.45ex width0.03em depth0em
          \kern-0.30em\rm C}}
\def\C{{\mathchoice
  {\sCC}
  {\sCC}
  {\kern 0.225em \vrule height1.05ex width0.025em depth0em \kern-0.25em \rm C}
  {\kern 0.180em \vrule height0.78ex width0.02em depth0em \kern-0.2em \rm C}
        }}
\def\sHH{{\rm I\kern-.16em{}H}}
\def\H{{\mathchoice
  {\sHH}
  {\sHH}
  {\rm I\kern-.13em{}H}
  {\rm I\kern-.13em{}H} }}
\def\sNN{{\rm I\kern-.16em{}N}}
\def\N{{\mathchoice
  {\sNN}
  {\sNN}
  {\rm I\kern-.12em{}N}
  {\rm I\kern-.10em{}N} }}
\def\sPP{{\rm I\kern-.16em{}P}}
\def\P{{\mathchoice
  {\sPP}
  {\sPP}
  {\rm I\kern-.12em{}P}
  {\rm I\kern-.10em{}P} }}
\def\sQQ{{\kern 0.27em \vrule height1.45ex width0.03em depth0em
          \kern-0.30em \rm Q}}
\def\Q{{\mathchoice
        {\sQQ}
        {\sQQ}
  {\kern 0.225em \vrule height1.05ex width0.025em depth0em \kern-0.25em \rm Q}
  {\kern 0.180em \vrule height0.78ex width0.020em depth0em \kern-0.20em \rm Q}
        }}
\def\sRR{{\rm I\kern-0.16em{}R}}
\def\R{{\mathchoice
  {\sRR}
  {\sRR}
  {\rm I\kern-0.12em{}R}
  {\rm I\kern-0.10em{}R} }}
\def\sZZ{{\rm Z\kern-0.32em{}Z}}
\def\Z{{\mathchoice
  {\sZZ}
  {\sZZ} 
  {\rm Z\kern-0.3em{}Z}     %.3
  {\rm Z\kern-0.25em{}Z} }}  %.25
\def\ZZZ{{\rm Z\kern-0.24em{}Z}}
\def\sII{{\rm I\kern-0.16em{}I}}
\def\I{{\mathchoice
  {\sII}
  {\sII}
  {\rm I\kern-0.12em{}I}
  {\rm I\kern-0.10em{}I} }}

\def\inbar{\,\vrule height1.5ex width.4pt depth0pt}
\font\cmss=cmss10 \font\cmsss=cmss10 at 7pt

\def\smiley{\hbox{\large$\bigcirc$\hspace{-0.80em}\raise.2ex
\hbox{$\cdot\cdot$}\kern-.61em\lower.2ex\hbox{\scriptsize$\smile$}}\ }
\def\frowny{\hbox{\large$\bigcirc$\hspace{-0.80em}\raise.2ex
\hbox{$\cdot\cdot$}\kern-.635em\lower.2ex\hbox{\scriptsize$\frown$}}\ }

\def\I{{\rlap{1} \hskip 1.6pt \hbox{1}}}

\makeatletter
\let\hangafter\@hangfrom
\makeatother

\newtheorem{definition}{\sf DEFINITION}

\preprint{MIT-CTP-2871\\ \\ {\tt hep-th/}}

\title{A Monograph on the Classification of the Discrete 
	Subgroups of $SU(4)$}
\author{Amihay Hanany and Yang-Hui He\footnote{
Research supported in part
by the CTP and LNS of MIT and the U.S. Department of 
Energy under cooperative research
agreement $\#$DE-FC02-94ER40818. YHH is also supported by the NSF Graduate
Fellowship.}
\\
Center for Theoretical Physics,
\\ Massachusetts Institute of Technology,\\ Cambridge, MA 02139, USA.\\
\email{hanany, yhe@ctp.mit.edu}
}
\abstract{We here present, in modern notation, the classification of 
the discrete finite
subgroups of $SU(4)$ as well as the
\href{http://pierre.mit.edu/~yhe/su4.ct}{character tables} 
for the exceptional cases
thereof (Cf. http://pierre.mit.edu/$\sim$yhe/su4.ct).
We hope this catalogue will be useful to works on string orbifold theories,
quiver theories, WZW modular invariants, Gorenstein resolutions, 
nonlinear sigma-models
as well as some recently proposed inter-connections among them.}
\begin{document}
\section{Introduction}
It is well known that the discrete finite subgroups of $SL(n=2,3;\C)$ have been
completely classified; works related to string orbifold theories and quiver
theories have of late used these results (see for example \cite{He1,He2,Orb,Muto} 
as well as
references therein). Conjectures regarding higher $n$ have been raised and works
toward finite subgroups of $SU(4)$ are under way.
Recent works by physicists and mathematicians alike further beckon for a 
classification of the groups, conveniently presented,
in the case of $SU(4)$ \cite{Vafa}.
Compounded thereupon is the disparity of language under which the groups are discussed:
the classification problem in the past decades has chiefly been of interest to either
theoretical chemists or to pure mathematicians, the former of whom disguise them in
Bravais crystallographic notation (e.g.\ \cite{Chem}) while the latter abstract them in
fields of finite characteristic (e.g.\ \cite{Math}). Subsequently, there is
a need within the string theory community for a list of the finite subgroups of
$SU(4)$ tabulated in our standard nomenclature, complete with the generators
and some brief but not overly-indulgent digression on their properties.

The motivations for this need are manifold. There has recently been a host
of four dimensional finite gauge theories constructed by placing D3 branes
on orbifold singularities \cite{Orb}; brane setups have also been
achieved for some of the groups \cite{HU}. 
In particular, a theory with ${\cal N}=2,1,0$
supercharges respectively
is obtained from a $\C^N/\{\Gamma \subset SU(n=2,3,4)\}$ singularity with $N=2,3$
(see \cite{He1} \cite{Orb} and references therein). Now as mentioned above $n=2,3$ have
been discussed, and $n=4$ has yet to be fully attacked. This last case is of particular
interest because it gives rise to an ${\cal N}=0$, non-supersymmetric
theory. On the one hand these orbifold theories provide interesting
string backgrounds for checks on the AdS/CFT Correspondence \cite{CFT}.
On the other hand, toric descriptions for the Abelian cases of the
canonical Gorenstein singularities have been treated while the non-Abelian
still remain elusive \cite{Agata}.
Moreover, the quiver theories arising from these string orbifold
theories (or equivalently, representation rings of finite subgroups of
$SU(n)$) have been hinted to be related to modular invariants of
$\widehat{su(n)}$-WZW models (or equivalently, affine characters of
$\widehat{su(n)}$) for arbitrary $n$ \cite{He1,DiFrancesco}, and a generalised
McKay Correspondence, which would also relate non-linear sigma models, 
 has been suggested to provide a reason \cite{He2}. Therefore a need for
the discrete subgroups of $SU(4)$ arises in all these areas.

Indeed the work has been done by Blichfeldt \cite{Blichfeldt} in 1917, or at least
 all the exceptional cases, though in an obviously outdated parlance and moreover with many   
infinite series being ``left to the reader as an exercise.'' It is therefore the
intent of the ensuing monograph to present the discrete subgroups $\Gamma$ of $SL(4,\C)$
in a concise fashion, hoping it to be of use to impending work, particularly 
non-supersymmetric conformal gauge theories from branes on orbifolds, resolution of
Gorenstein singularities in higher dimension, as well as
$\widehat{su(4)}$-WZW models.

\section*{Nomenclature}
Unless otherwise stated we shall adhere to the convention that
$\Gamma$ refers to a discrete subgroup of $SU(n)$ (i.e., a finite collineation group),
that $<x_1,..,x_n>$ is a finite group generated by $\{x_1,..,x_n\}$, that
$H \triangleleft G$ means $H$ is a normal subgroup of $G$, that
$S_n$ and $A_n$ are respectively the symmetric and alternating permutation groups on
$n$ elements, and that placing $*$ next to a group signifies that it belongs to
$SU(4) \subset SL(4;\C)$.

\section{Preliminary Definitions}
Let $\Gamma$ be a finite discrete subgroup of the general linear group, i.e.,
$\Gamma \subset GL(n,\C)$.
From a mathematical perspective, quotient varieties of the
form $\C^n / \Gamma$ may be constructed and by the theorem of Khinich and Watanabe
\cite{Khinich,Yau}, the quotient is Gorenstein\footnote{That is, if there 
exists a nowhere-vanishing holomorphic $n$-form. These varieties thus provide
local models of Calabi-Yau manifolds and are recently of great interest.}
if and only if 
$\Gamma$ is in fact in $SL(n,\C)$. Therefore we would like to focus on the
discrete subgroups of linear transformations {\it up to linear equivalence}, 
which are what has been dubbed in the old literature as 
{\bf finite collineation groups} \cite{Blichfeldt}. 
From a physics perspective, discrete subgroups of $SU(n) \subset SL(n;\C)$
have been subject to investigation in the early days of particle phenomenology 
\cite{Klink} and have lately been of renewed interest in string theory, 
especially in the context of orbifolds (see for example \cite{He1,He2,Orb,Vafa}).

There are some standard categorisations of finite collineation 
groups \cite{Blichfeldt,Yau}. They
first fall under the division of transitivity and intransitivity as follows:
\begin{definition}
If the $n$ variables upon which $\Gamma$ acts as a linear transformation
can be separated into 2 or more sets either directly or after a change of
variables, such that the variables of each set are transformed into linear
functions only of themselves, then $\Gamma$ is called {\bf Intransitive}; it is
called Transitive otherwise.
\end{definition}

The transitive $\Gamma$ can be further divided into the primitive and 
imprimitive cases:
\begin{definition}
If for the transitive $\Gamma$ the variables may be separated\footnote{
Again, either directly or after a change of variables.} into 2 or more
sets such that the variables of each are transformed into linear functions
of only those in any set according to the separation (either the same or
different), then $\Gamma$ is called {\bf Imprimitive}; it is called 
Primitive otherwise.
\end{definition}

Therefore in the  matrix representation of the groups, 
we may na\"{\i}vely construe
intransitivity as being block-diagonalisable and imprimitivity as being
block off-diagonalisable, whereby making primitive groups generically having
no {\it a priori} zero entries. We give examples of an intransitive,
a (transitive) imprimitive and a (transitive) primitive group, 
in their matrix forms, as follows:
\[
\begin{array}{ccc}
\left(\matrix{	\times &  \times &  0 &  0\\
		\times &  \times &  0 &  0 \\ 
		0 &  0 &  \times &  \times \\
		0 &  0 &  \times &  \times}
\right)
&
\left(\matrix{	0 &  0 &  \times &  \times \\
		0 &  0 &  \times &  \times \\
		\times &  \times &  0 &  0 \\
		\times &  \times &  0 &  0}
\right)
&
\left(\matrix{	\times &  \times &  \times &  \times \\
		\times &  \times &  \times &  \times \\
		\times &  \times &  \times &  \times \\
		\times &  \times &  \times &  \times}
\right)
\\
$Intransitive$	&	$Imprimitive$ 	& $Primitive$ \\
& \multicolumn{2}{c}{$Transitive$} \\
\end{array}
\]
Let us diagrammatically summarise all these inter-relations as is done in \cite{Yau}:
\[
\Gamma \left\{\begin{array}{l}
	$Intransitive$ \\
	$Transitive$ \left\{\begin{array}{l}
		$Imprimitive$\\
		$Primitive$ \left\{\begin{array}{l}
			$Simple$ \\
			$Having Normal Primitive Subgroups$ \\
			$Having Normal Intransitive Subgroups$ \\
			$Having Normal Imprimitive Subgroups$ \\
		\end{array} \right.
	\end{array}  \right. 
\end{array}  \right.
\]

In some sense the primitive groups are the fundamental building 
blocks and
pose as the most difficult to be classified. It is those primitive groups
that Blichfeldt presented, as linear transformations, in \cite{Blichfeldt}.
These groups are what we might call {\it exceptionals} in the sense that they 
do not fall into infinite series, in analogy to the $E_{6,7,8}$ groups of
$SU(2)$. We present them as well as their sub-classifications first. Thereafter
we shall list the imprimitive and intransitives, which give rise to a host of
infinite series of groups, in analogy to the $A_n$ and $D_n$ of $SU(2)$.

Let us take a final digression to clarify the so-called 
{\bf Jordan Notation}, which is the
symbol $\phi$ commonly used in finite group theory. A linear
group $\Gamma$ often has its order
denoted as $|\Gamma| = g\phi$ for positive integers $g$ and $\phi$; 
the $\phi$ signifies the order of the subgroup of homotheties, or those
multiples of the identity which together form the center of the $SL(n;\C)$.
We know that $SU(n) \subset SL(n;\C)$, so a subgroup of the latter is not
necessarily that of the former. In the case of $SL(n=2,3;\C)$,
the situation is
simple\footnote{See \cite{Klink,He1} for a discussion on this point.}:
the finite subgroups belonged either to (A) $SU(n=2,3)$, or to (B) the 
center-modded\footnote{For $n=2$, this our familiar $SU(2)/\Z_2 \cong SO(3)$.}
$SU(n=2,3)/\Z_{2,3}$, or (C) to both. Of course a group with order $g$ 
in type (B) would have a natural lifting to type (A) and become a group
of order $g$ multiplied by $|\Z_2|=2$ or $|\Z_3=3|$ respectively, 
which is now a finite subgroup of the full $SU(2)$ or $SU(3)$, implying that
the Jordan $\phi$ is 2 or 3 respectively.

For the case at hand, the situation is slightly more complicated since 4 is
not a prime. Therefore $\phi$ can be either 2 or 4 depending how one lifts
with respect to the relation $SU(4)/\Z_2\times\Z_2 \cong SO(6)$ and we lose
a good discriminant of whether or not $\Gamma$ is in the full $SU(4)$. To
this end we have explicitly verified the unitarity condition for the group
elements and will place a star ($*$) next to those following groups which indeed
are in the full $SU(4)$. Moreover, from the viewpoint of string orbifold
theories which study for example the fermionic and bosonic matter content
of the resulting Yang-Mills theory, one naturally takes interest in
$Spin(6)$, or the full $\Z_2 \times \Z_2$ cover of $SO(6)$ which admits
spinor representations; for these we shall
look in particular at the groups that have $\phi = 4$ in the Jordan notation,
as will be indicated in the tables below.

\section{The Discrete Finite Subgroups of $SL(4;\C)$}
We shall henceforth let $\Gamma$ denote a finite subgroup of $SL(4;\C)$
unless otherwise stated.
\subsection{Primitive Subgroups}
There are in all 30 types of primitive cases for $\Gamma$. 
First we define the constants $w = e^{\frac{2\pi i}{3}}$, 
$\beta = e^{\frac{2\pi i}{7}}$,
$p = \beta + \beta^2 +\beta^4$, $q = \beta^3 + \beta^5 +\beta^6$,
$s = \beta^2 + \beta^5$, $t = \beta^3 + \beta^4$, and $u = \beta + \beta^6.$ 
Furthermore
we shall adhere to some standard notation and denote the permutation 
and the alternating permutation group on $n$ elements 
respectively as $S_n$ and $A_n$. Moreover, in what follows we shall use the
function $Lift$ to mean the lifting by (perhaps a subgroup) of the Abelian center
$C$ according to the exact sequence
$
\begin{array}{ccccccccc}
0 & \rightarrow & C & \rightarrow & SU(4) & \rightarrow & SU(4)/C & \rightarrow
& 0. \\
\end{array}
$

We present the relevant matrix generators as we proceed:
{\small
\[
\begin{array}{c}
F_1 = \left(
\matrix{1 & 0 & 0 & 0 \\ 
	0 & 1 & 0 & 0 \\
	0 & 0 & w & 0 \\
	0 & 0 & 0 & w^2 \\}
\right)

F_2 = \frac{1}{\sqrt{3}}\left(
\matrix{1 & 0 & 0 & \sqrt{2} \\ 
	0 & -1 & \sqrt{2} & 0 \\
	0 & \sqrt{2} & 1 & 0 \\
	\sqrt{2} & 0 & 0 & -1 \\}
\right)

F_3 = \left(
\matrix{\frac{\sqrt{3}}{2} & \frac{1}{2} & 0 & 0 \\ 
	\frac{1}{2} & -\frac{\sqrt{3}}{2} & 0 & 0 \\
	0 & 0 & 0 & 1 \\
	0 & 0 & 1 & 0 \\}
\right)

\\ \\

F'_2 = \frac{1}{3}\left(
\matrix{3 & 0 & 0 & 0 \\ 
	0 & -1 & 2 & 2 \\
	0 & 2 & -1 & 2 \\
	0 & 2 & 2 & -1 \\}
\right)

F'_3 = \frac{1}{4}\left(
\matrix{-1 & \sqrt{15} & 0 & 0 \\ 
	\sqrt{15} & 1 & 0 & 0 \\
	0 & 0 & 0 & 4 \\
	0 & 0 & 4 & 0 \\}
\right)

F_4 = \left(
\matrix{0 & 1 & 0 & 0 \\ 
	1 & 0 & 0 & 0 \\
	0 & 0 & 0 & -1 \\
	0 & 0 & -1 & 0 \\}
\right)

\\ \\

S = \left(
\matrix{1 & 0 & 0 & 0 \\ 
	0 & \beta & 0 & 0 \\
	0 & 0 & \beta^4 & 0 \\
	0 & 0 & 0 & \beta^2 \\}
\right)

T = \left(
\matrix{1 & 0 & 0 & 0 \\ 
	0 & 0 & 1 & 0 \\
	0 & 0 & 0 & 1 \\
	0 & 1 & 0 & 0 \\}
\right)

W = \frac{1}{i\sqrt{7}}\left(
\matrix{p^2 & 1 & 1 & 1 \\ 
	1 & -q & -p & -p \\
	1 & -p & -q & -p \\
	1 & -p & -p & -q \\}
\right)

\\ \\

R = \frac{1}{\sqrt{7}}\left(
\matrix{1 & 1 & 1 & 1 \\ 
	2 & s & t & u \\
	2 & t & u & s \\
	2 & u & s & t \\}
\right)

C = \left(
\matrix{1 & 0 & 0 & 0 \\ 
	0 & 1 & 0 & 0 \\
	0 & 0 & w & 0 \\
	0 & 0 & 0 & w^2 \\}
\right)

D = \left(
\matrix{w & 0 & 0 & 0 \\ 
	0 & w & 0 & 0 \\
	0 & 0 & w & 0 \\
	0 & 0 & 0 & 1 \\}
\right)

\\ \\

V = \frac{1}{i\sqrt{3}}\left(
\matrix{i\sqrt{3} & 0 & 0 & 0 \\ 
	0 & 1 & 1 & 1 \\
	0 & 1 & w & w^2 \\
	0 & 1 & w^2 & w \\}
\right)

F = \left(
\matrix{0 & 0 & -1 & 0 \\ 
	0 & 1 & 0 & 0 \\
	-1 & 0 & 0 & 0 \\
	0 & 0 & 0 & -1 \\}
\right)

\end{array}
\]
}
We see that all these matrix generators are unitary except $R$.

\subsubsection{Primitive Simple Groups}
There are 6 groups of this most fundamental type:
\[
\begin{array}{|c|c|c|c|}
\hline
$Group$	& $Order$	& $Generators$ 	& $Remarks$ \\
\hline \hline
$I$*	& 60\times 4	& F_1,F_2,F_3	& Lift(A_5)\\
$II$*	& 60		& F_1,F'_2,F'_3 & \cong A_5 \\
$III$*	& 360\times 4	& F_1,F_2,F_3	& Lift(A_6)\\
$IV$*	& \frac12 7!\times 2	& S,T,W		& Lift(A_7)\\
$V$	& 168\times 4	& S,T,R		& \\
$VI$*	& 2^6 3^4 5\times 2& T,C,D,E,F	& \\	
\hline
\end{array}
\]

\subsubsection{Groups Having Simple Normal Primitive Subgroups}
There are 3 such groups, generated by simple primitives
and the following 2 matrices:

{\small
\[
\begin{array}{c}
F' = \frac{1+i}{\sqrt{2}}\left(
\matrix{1 & 0 & 0 & 0 \\ 
	0 & 1 & 0 & 0 \\
	0 & 0 & 0 & 1 \\
	0 & 0 & 1 & 0 \\}
\right)

F'' = \left(
\matrix{0 & 1 & 0 & 0 \\ 
	-1 & 0 & 0 & 0 \\
	0 & 0 & 0 & 1 \\
	0 & 0 & -1 & 0 \\}
\right)
\end{array}
\]
}
The groups are then:
\[
\begin{array}{|c|c|c|c|}
\hline
$Group$	& $Order$	& $Generators$ 	& $Remarks$ \\
\hline \hline
$VII$*	& 120\times 4	& ($I$),F''	& Lift(S_5)\\
$VIII$*	& 120\times 4	& ($II$),F'	& Lift(S_5)\\
$IX$*	& 720\times 4	& ($III$),F''	& Lift(S_6)\\
\hline
\end{array}
\]

\subsubsection{Groups Having Normal Intransitive Subgroups}
There are seven types of $\Gamma$ in this case and their fundamental
representation matrices turn out to be Kronecker products of 
those of the exceptionals of $SU(2)$. In other words, 
for $M$, the matrix representation
of $\Gamma$, we have $M = A_1 \otimes_K A_2$ such that $A_i$ are the 
$2\times 2$ matrices representing $E_{6,7,8}$.
Indeed we know that $E_6 = \langle S_{SU(2)}, U^2_{SU(2)}\rangle,
E_7 = \langle S_{SU(2)}, U_{SU(2)}\rangle,
E_8 = \langle S_{SU(2)}, U^2_{SU(2)}, V_{SU(2)} \rangle$, where
{\small
\[
\begin{array}{c}
S_{SU(2)} = \frac12\left(
\matrix{-1 + i	& -1 + i \\
	1 + i	& -1 - i \\}
\right)

U_{SU(2)} = \frac{1}{\sqrt{2}}\left(
\matrix{1 + i	& 0 \\
	0	& 1 - i \\}
\right).

\\ \\

V_{SU(2)} = \left(
\matrix{\frac{i}{2}	& \frac{1-\sqrt{5}}{4} - i \frac{1+\sqrt{5}}{4}\\
	-\frac{1-\sqrt{5}}{4} - i \frac{1+\sqrt{5}}{4}	& -\frac{i}{2} \\}
\right)
\end{array}
\]
}

We use, for the generators, the notation $\langle A_i\rangle \otimes \langle B_j \rangle$
to mean that Kronecker products are to be formed between all combinations of $A_i$ with
$B_j$. Moreover the group (XI), a normal subgroup of (XIV), is formed by tensoring the
2-by-2 matrices 
$x_1 = \frac{1}{\sqrt{2}}\left(\matrix{1 & 1\cr i & -i}\right)$,
$x_2 = \frac{1}{\sqrt{2}}\left(\matrix{i & i\cr -1 & 1}\right)$,
$x_3 = \frac{1}{\sqrt{2}}\left(\matrix{-1 & -1\cr -1 & 1}\right)$,
$x_4 = \frac{1}{\sqrt{2}}\left(\matrix{i & 1\cr 1 & i}\right)$,
$x_5 = \frac{1}{\sqrt{2}}\left(\matrix{1 & -1\cr -i & -i}\right)$, and
$x_6 = \frac{1}{\sqrt{2}}\left(\matrix{i & -i\cr 1 & 1}\right)$.
The seven groups are:
\[
\begin{array}{|c|c|c|c|}
\hline
$Group$	& $Order$	& $Generators$ 	& $Remarks$ \\
\hline \hline
$X$*	& 144 \times 2	
	& \langle S_{SU(2)}, U^2_{SU(2)}\rangle \otimes \langle S_{SU(2)}, U^2_{SU(2)}\rangle
	& \cong E_6 \otimes_K E_6\\
$XI$*	& 288 \times 2
	& x_1 \otimes x_2, x_1 \otimes x^T_2, x_3 \otimes x_4, x_5 \otimes x_6
	& ($X$) \triangleleft \Gamma \triangleleft ($XIV$)\\
$XII$*	& 288 \times 2	
	& \langle S_{SU(2)}, U^2_{SU(2)}\rangle \otimes \langle S_{SU(2)}, U_{SU(2)}\rangle
	& \cong E_6 \otimes_K E_7\\
$XIII$*	& 720 \times 2
	& \langle S_{SU(2)}, U^2_{SU(2)}\rangle 
		\otimes \langle S_{SU(2)}, V_{SU(2)}, U^2_{SU(2)}\rangle		
	& \cong E_6 \otimes_K E_8\\
$XIV$*	& 576 \times 2	
	& \langle S_{SU(2)}, U_{SU(2)}\rangle \otimes \langle S_{SU(2)}, U_{SU(2)}\rangle
	& \cong E_7 \otimes_K E_7\\
$XV$*	& 1440 \times 2	
	& \langle S_{SU(2)}, U_{SU(2)}\rangle \otimes \langle S_{SU(2)}, V_{SU(2)}, U^2_{SU(2)}\rangle	
	& \cong E_7 \otimes_K E_8\\
$XVI$*	& 3600 \times 2	
	& \langle S_{SU(2)}, V_{SU(2)}, U^2_{SU(2)}\rangle 
		\otimes \langle S_{SU(2)}, V_{SU(2)}, U^2_{SU(2)}\rangle
	& \cong E_8 \otimes_K E_8\\
\hline
\end{array}
\]

\subsubsection{Groups Having X-XVI as Normal Primitive Subgroups}
There are in all 5 of these, generated by the above, together with
{\small
\[
\begin{array}{c}
T_1 = \frac{1+i}{\sqrt{2}}\left(
\matrix{1 & 0 & 0 & 0 \\ 
	0 & 0 & 1 & 0 \\
	0 & 1 & 0 & 0 \\
	0 & 0 & 0 & 1 \\}
\right)

T_2 = \left(
\matrix{1 & 0 & 0 & 0 \\ 
	0 & 0 & 1 & 0 \\
	0 & i & 0 & 0 \\
	0 & 0 & 0 & i \\}
\right)
\end{array}
\]
}
The group generated by (XIV) and $T_2$ is isomorphic to (XXI), generated by
(XIV) and $T_1$ so we need not consider it. The groups are:
\[
\begin{array}{|c|c|c|}
\hline
$Group$	& $Order$	& $Generators$ 	\\
\hline \hline
$XVII$*	&576\times 4 	& ($XI$),T_1 	\\
$XVIII$*&576\times 4  	& ($XI$),T_2	\\
$XIX$*	&288\times 4 	& ($X$),T_1	\\
$XX$*	&7200\times 4 	& ($XVI$),T_1	\\
$XXI$*	&1152\times 4	& ($XIV$),T_1	\\
\hline
\end{array}
\]

\subsubsection{Groups Having Normal Imprimitive Subgroups}
Finally these following 9 groups of order divisible by 5
complete our list of the primitive
$\Gamma$, for which we need the following generators:
{\small
\[
\begin{array}{c}
A = \frac{1+i}{\sqrt{2}}\left(
\matrix{1 & 0 & 0 & 0 \\ 
	0 & i & 0 & 0 \\
	0 & 0 & i & 0 \\
	0 & 0 & 0 & 1 \\}
\right)

B = \frac{1+i}{\sqrt{2}}\left(
\matrix{1 & 0 & 0 & 0 \\ 
	0 & 1 & 0 & 0 \\
	0 & 0 & 1 & 0 \\
	0 & 0 & 0 & -1 \\}
\right)

\\ \\

S' = \frac{1+i}{\sqrt{2}}\left(
\matrix{i & 0 & 0 & 0 \\ 
	0 & i & 0 & 0 \\
	0 & 0 & 1 & 0 \\
	0 & 0 & 0 & 1 \\}
\right)

T' = \frac{1+i}{2}\left(
\matrix{-i & 0 & 0 & i \\ 
	0 & 1 & 1 & 0 \\
	1 & 0 & 0 & 1 \\
	0 & -i & i & 0 \\}
\right)

R' = \frac{1}{\sqrt{2}}\left(
\matrix{1 & i & 0 & 0 \\ 
	i & 1 & 0 & 0 \\
	0 & 0 & i & 1 \\
	0 & 0 & -1 & -i \\}
\right)

\end{array}
\]
}

Moreover these following groups contain the group $K$ of order
$16\times 2$, generated by:
{\small
\[
\begin{array}{c}
A_1 = \left(
\matrix{1 & 0 & 0 & 0 \\ 
	0 & 1 & 0 & 0 \\
	0 & 0 & -1 & 0 \\
	0 & 0 & 0 & -1 \\}
\right)

A_2 = \left(
\matrix{1 & 0 & 0 & 0 \\ 
	0 & -1 & 0 & 0 \\
	0 & 0 & -1 & 0 \\
	0 & 0 & 0 & 1 \\}
\right)

\\ \\

A_3 = \left(
\matrix{0 & 1 & 0 & 0 \\ 
	1 & 0 & 0 & 0 \\
	0 & 0 & 0 & 1 \\
	0 & 0 & 1 & 0 \\}
\right)

A_4 = \left(
\matrix{0 & 0 & 1 & 0 \\ 
	0 & 0 & 0 & 1 \\
	1 & 0 & 0 & 0 \\
	0 & 1 & 0 & 0 \\}
\right)

\end{array}
\]
}
We tabulate the nine groups:
\[
\begin{array}{|c|c|c|}
\hline
$Group$	& $Order$		& $Generators$ 	\\
\hline \hline
$XXII$*	& 5\times 16 \times 4	& ($K$),T'	\\
$XXIII$*& 10\times 16 \times 4	& ($K$),T',R'^2	\\
$XXIV$*	& 20\times 16 \times 4	& ($K$),T,R	\\
$XXV$*	& 60\times 16 \times 4	& ($K$),T,S'B	\\
$XXVI$*	& 60\times 16 \times 4	& ($K$),T,BR'	\\
$XXVII$*& 120\times 16 \times 4	& ($K$),T,A	\\
$XXVIII$*& 120\times 16 \times 4& ($K$),T,B	\\
$XXIX$*	& 360\times 16 \times 4	& ($K$),T,AB	\\
$XXX$*	& 720\times 16 \times 4	& ($K$),T,S	\\
\hline
\end{array}
\]

\subsection{Intransitive Subgroups}
These cases are what could be constructed from the various combinations of the
discrete subgroups of $SL(2;\C)$ and $SL(3;\C)$ according to the various
possibilities of diagonal embeddings. Namely, they consist of those of the
form $(1,1,1,1)$ which represents the various possible Abelian groups with 
one-dimensional (cyclotomic) representation\footnote{These includes the
$\Z_m \times \Z_n \times \Z_p$ groups recently of interest in brane cube
constructions \cite{Uranga}.}, $(1,1,2)$, two Abelians and an $SL(2;\C)$ subgroup,
$(1,3)$, an Abelian and an $SL(3;\C)$ subgroup, and $(2,2)$, two $SL(2;\C)$ subgroups
as well as the various permutations thereupon. Since these embedded groups (as
collineation groups of lower dimension) have been well discussed \cite{He1}, we
shall not delve too far into their account.

\subsection{Imprimitive Groups}
The analogues of the dihedral groups (in both $SL(2;\C)$ and $SL(3;\C)$), which
present themselves as infinite series, are to be found in these last cases of 
$\Gamma$. They are of two subtypes:
\begin{itemize}
\item (a) Generated by the canonical Abelian group of order $n^3$ for $n \in \Z^+$
	whose elements are
	\[
	\begin{array}{cc}
	\Delta =  \{\left(
	\matrix{\omega^i & 0 & 0 & 0 \\ 
		0 & \omega^j & 0 & 0 \\
		0 & 0 & \omega^k & 0 \\
		0 & 0 & 0 & \omega^{-i-j-k} \\}
	\right)\}
	&
	\begin{array}{c}
		\omega = e^{\frac{2 \pi i}{n}} \\
		i,j,k = 1,...,n
	\end{array}
	\end{array}
	\]
	as well as respectively the four groups $A_4$, $S_4$, the Sylow-8 subgroup
	$Sy \subset S_4$ (or the ordinary dihedral group of 8 elements)
	 and $\Z_2 \times \Z_2$;
\item (b) We define $H$ and $T''$ (where again $i = 1,...,n$) as:
	\[
	\begin{array}{c}
	H = \left(
	\matrix{a & b & 0 & 0 \\ 
		c & d & 0 & 0 \\
		0 & 0 & e & f \\
		0 & 0 & g & h \\}
	\right)

	T'' = \left(
	\matrix{0 & 0 & 1 & 0 \\ 
		0 & 0 & 0 & 1 \\
		\omega^i & 0 & 0 & 0 \\
		0 & \omega^{-i} & 0 & 0 \\}
	\right)
	\end{array}
	\]
	where the blocks of $H$ are $SL(2;\C)$ subgroups.
\end{itemize}
We tabulate these last cases of $\Gamma$ as follows:
\[
\begin{array}{|c|c|c|c|}
\hline
$Subtype$	& $Group$	& $Order$ & $Generators$ \\
\hline \hline
(a)		& $XXXI$*	& 12n^3	& \langle \Delta,A_4 \rangle\\
		& $XXXII$*	& 24n^3	& \langle \Delta,S_4 \rangle\\
		& $XXXIII$*	& 8n^3	& \langle \Delta,Sy \rangle\\
		& $XXXIII$*	& 4n^3	& \langle \Delta,\Z_2 \times \Z_2 \rangle\\
(b)		& $XXXIV$*	& 	& \langle H,T'' \rangle \\
\hline
\end{array}
\]

\section{Remarks}
We have presented, in modern notation, the classification of the discrete subgroups
of $SL(4,\C)$ and in particular, of $SU(4)$. The matrix generators and orders of these
groups have been tabulated, while bearing in mind how the latter fall into 
sub-categories of transitivity and primitivity standard to discussions on collineation
groups.

Furthermore, we have computed the character table for the 30 exceptional 
cases \cite{GAP}; 
The interested reader may, at his or her convenience, find the character tables
at 
\href{http://pierre.mit.edu/~yhe/su4.ct}{http://pierre.mit.edu/$\sim$yhe/su4.ct}.
These tables will
be crucial to quiver theories.
As an example, we present in \fref{fig:I} the quiver for the irreducible {\bf 4} of
the group (I) of
order $60 \times 4$, which is the lift of the alternating 
permutation group on 5 elements.

%%%%%%%%%%%%%%%%%%%%%%%%%%%%%%%%%%%%%%%%%%%%%%%%%%%%%%%
\EPSFIGURE[ht]{I.eps,width=7in}
{
\label{fig:I}
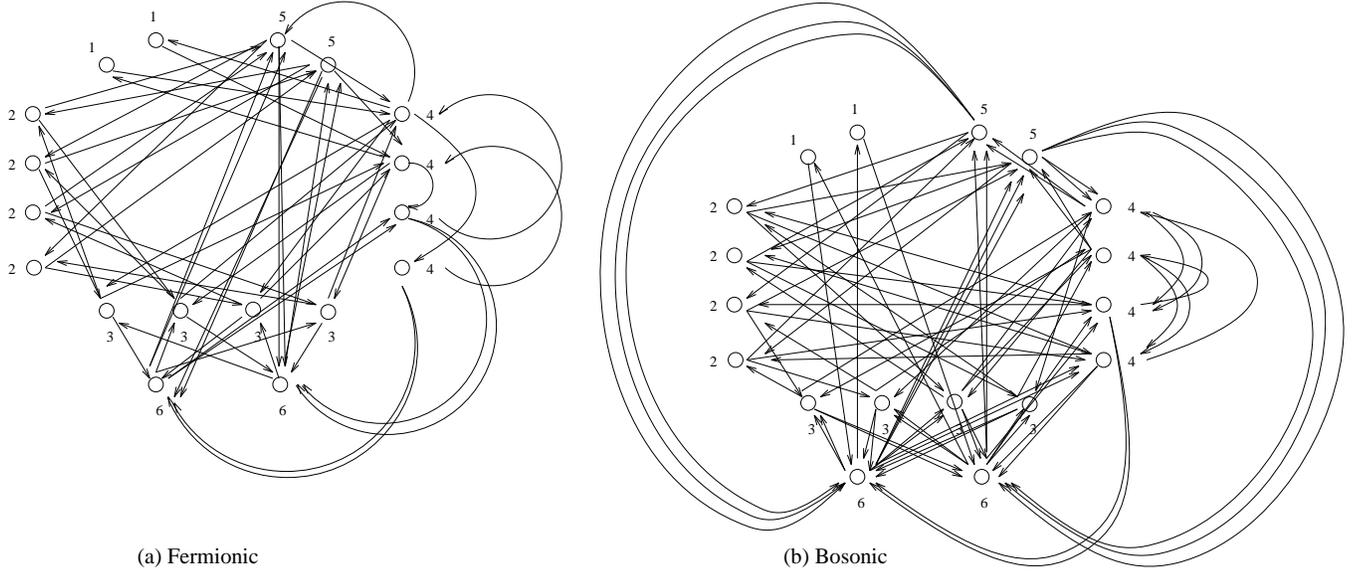
The Quiver Diagram for Group (I), constructed for (a) the fermionic
$a_{ij}^4$ corresponding to the irreducible ${\bf 4}_3$ and (b) the bosonic
$a_{ij}^6$ corresponding to the irreducible ${\bf 6}_2$  (in the notation
of \cite{He1}). We make this choice because we know that 
${\bf 4}_1 \otimes {\bf 4}_3 = {\bf 4}_3 \oplus {\bf 6}_1 \oplus {\bf 6}_2$
and that the two {\bf 6}'s are conjugates. 
The indices are the 
dimensions of the various
irreducible representations, a generalisation of Dynkin labels.
}
%%%%%%%%%%%%%%%%%%%%%%%%%%%%%%%%%%%%%%%%%%%%%%%%%%%%%%%

Indeed such quiver diagrams may be constructed for all the groups using
the character tables mentioned above. We note in passing that since
$\Gamma \subset SU(4)$ gives rise to an ${\cal N}=0$ theory in 4 dimensions,
supersymmetry will not come to our aid in relating the fermionic $a_{ij}^{\bf 4}$
and the bosonic $a_{ij}^{\bf 6}$ as was done in \cite{He1}.
However we can analyse the problem with a slight modification and place
a stack of M2 branes on the orbifold,
(which in the Maldacena picture corresponds to orbifolds on the $S^7$ factor
in $AdS_4 \times S^7$), 
and obtain an ${\cal N}=2$ theory in 3 dimensions at least in the
IR limit as we lift from type IIA to M Theory \cite{Orb,Agata,Fabbri}.
This supersymmetry would help us to impose the constraining relation between
the two matter matrices, and hence the two quiver diagrams. This would be
an interesting check which we leave to future work.

We see therefore a host of prospective research in various areas,
particularly in the context of string orbifold/gauge 
theories, WZW modular invariants, and singularity-resolutions in algebraic geometry.
It is hoped that this monograph, together with its companion tables on the web,
will provide a ready-reference to works in these directions.

\section*{Acknowledgements}
{\it Ad Catharinae Sanctae Alexandriae et Ad Majorem Dei Gloriam...\\}
We would like to express our sincere gratitude to B. Feng for tireless discussions, 
as well as R. Britto-Pacumio, K. McGerty, L. Ng, J. S. Song, 
M. B. Spradlin and A. Uranga for valuable comments.
YHH would also like to thank his parents, D. Matheu, I. Savonije and the Schmidts 
(particularly L. A. Schmidt) for their constant emotional support as well as
the CTP and the NSF for their gracious patronage.

\end{document}